\documentclass[epsf,useAMS,usenatbib, usegraphicx]{mn2e}
\usepackage{graphicx}
\usepackage{epsfig}
\usepackage{amsmath}
\usepackage{amssymb}
\usepackage{rotating}
\usepackage{color}
\usepackage{pifont}
\usepackage{longtable}
\usepackage{pdflscape,lscape}
\usepackage{float}
\usepackage{placeins}

\newcommand{\teff}{$T_{\rm eff}$}
\newcommand{\logg}{$\log g$}
\newcommand{\vsini}{$v \sin i$}
\newcommand{\kms}{km\,s$^{-1}$}
\newcommand{\ds}{$\delta$ Scuti}
\newcommand{\vr}{$v$$_{r}$}

\usepackage{appendix}
\usepackage{booktabs}
\usepackage{upgreek}
\usepackage{url}

\usepackage{natbib}
\title[TZ\,Dra]{Mass transfer and tidally tilted pulsation
in the Algol-type system TZ\,Dra}

\author[F. Kahraman Ali\c{c}avu\c{s} et. al.]{F. Kahraman Ali\c{c}avu\c{s}$^{1,2}$\thanks{E-mail: filizkahraman01@gmail.com}, G. Handler$^{3}$, F. Ali\c{c}avu\c{s}$^{1,2}$, P. De Cat$^{4}$, T. R. Bedding$^{5,6}$, \and P. Lampens$^{4}$, \"{O}. Ekinci$^{1,2}$, D. G\"{u}m\"{u}\textcommabelow{s}$^{7}$, F. Leone$^{8,9}$ 
\\
$^{1}$\c{C}anakkale Onsekiz Mart University, Faculty of Sciences and Arts, Physics Department, 17100, Canakkale, Turkey\\
$^{1}$\c{C}anakkale Onsekiz Mart University, Astrophysics Research Center and Ulup{\i}nar Observatory, TR-17100, Çanakkale, Turkey\\
$^{3}$Nicolaus Copernicus Astronomical Center, Polish Academy of Sciences, Bartycka 18, PL-00-716 Warsaw, Poland\\ 
$^{4}$Royal Observatory of Belgium, Ringlaan 3, B-1180 Brussel, Belgium\\
$^{5}$Sydney Institute for Astronomy (SIfA), School of Physics, University of Sydney, Camperdown NSW 2006, Australia\\
$^{6}$Stellar Astrophysics Centre, Department of Physics and Astronomy, Aarhus University, Ny Munkegade 120, DK-8000 Aarhus C,Denmark\\
$^{7}$Istanbul  University,  Institute  of  Graduate  Studies  in  Science,Programme  of  Astronomy  and  Space  Sciences,  34116,  Beyazıt,Istanbul, Turkey\\
$^{8}$Dipartimento di Fisica e Astronomia, Sezione Astrofisica, Universit´a di Catania, Via S. Sofia 78, I-95123 Catania, Italy\\
$^{9}$INAF, Osservatorio Astrofisico di Catania, Via S. Sofia 78, I-95123 Catania, Italy\\
}
\begin{document}

\date{Accepted ... Received ...; in original form ...}

\pagerange{\pageref{firstpage}--\pageref{lastpage}} \pubyear{2021}

\maketitle

\label{firstpage}

\begin{abstract}

 Oscillating eclipsing Algols (oEAs) are remarkable systems which allow us to determine accurate fundamental stellar parameters (mass, radius) and probe the stellar interiors through pulsations. TZ\,Dra is an oEA system containing a \ds\, component. To examine particular characteristics of such close systems including pulsations and mass transfer, we present a detailed photometric and spectroscopic study of TZ\,Dra. With the analysis of high-resolution spectra, the orbital parameters were determined by the radial velocity analysis and the atmospheric parameters were derived for the primary component. The binary modelling and the pulsational frequency analysis was carried out using the TESS data set. The H$\alpha$ line profiles show the signature of mass transfer from the cool to the hot binary component. The conclusion of mass transfer/mass loss in the system was supported by the analysis of the orbital period changes. As a result, it was found that there is $3.52 \times 10^{-9}$ $M_\odot$/year mass loss from the system most probably through the hotspot and stellar winds. Additionally, most pulsation frequencies originating from the primary component were found to be spaced by harmonics of the orbital frequencies in particular, twelve doublets spaced by $2f_{\rm orb}$ were detected from which we infer that this star a tidally tilted pulsator. A mean p-mode frequency spacing of $\approx 7.2$\,d$^{-1}$ was found as well.

\end{abstract}


\section{Introduction}

A significant fraction ($\sim$\,70\%) of stars in our galaxy is assumed to be a member of binary or multiple systems \citep{2011IAUS..272..474S, 2014MNRAS.443.3022A, 2017ApJS..230...15M}. A special group of binary systems is the eclipsing binaries. 
As the eclipsing binary stars are particular tools for the accurate determination of the fundamental stellar parameters such as mass ($M$) and radius ($R$), they are among the most studied stellar systems. 

There are also some extremely spectacular eclipsing binaries that consist of at least one pulsating star. These kinds of systems allows us to probe the stellar interiors through the study of pulsation frequencies occurring in the inner part of the oscillating binary component(s). Obtaining precise fundamental stellar parameters and having information about the stellar interior are certainly important and provides us powerful tools to check the validity of present stellar structure models and compare evolutionary models for single stars and stars in binary systems. Therefore, most studies have been carried out about these objects for decades \citep[e.g.][]{1978A&A....66..377J, 1984A&A...138..443B, 2003AJ....126.1933S, 2014A&A...565A..55D, 2021A&A...645A.119S}.  

Different kinds of high and intermediate-mass main-sequence oscillating stars like $\beta$ Cephei, $\delta$ Scuti and $\gamma$ Doradus can be components of eclipsing binaries \citep{2021Galax...9...28L, 2021arXiv211003543S}. However, the number of eclipsing binaries with \ds\, components is the largest compared to other groups of main sequence pulsating stars in eclipsing binaries \citep{2017MNRAS.470..915K, 2017MNRAS.465.1181L} because of their relatively shorter pulsation period and lower oscillation amplitude they are easier to detect. The \ds\, stars are $A-F$ type, dwarf to giant, pulsating variables which exhibit pulsations with frequencies higher than 5 d$^{-1}$ \citep{2000ASPC..210....3B, 2011MNRAS.417..591B, 2018MNRAS.476.3169B, 2019MNRAS.485.2380M}. Their oscillations are thought to be excited by the $\kappa$ mechanism operating in the He ionization zone \citep{1962ZA.....54..114B}. The \ds\, stars are located in the lower part of the classical Cepheid instability strip and they show pulsations in pressure (p) and gravity (g) modes which give information about different parts of the stellar interior.  

The first discoveries of \ds\, stars in eclipsing binary systems were presented in the beginning of the 1970s \citep{1971IBVS..596....1T}. Since that time, the number of known \ds\, stars in eclipsing binaries has been increased. Significant attempts to find new candidates of this kind of systems were done by \cite{2002ASPC..256..259M, 2005ASPC..333..197M} and \cite{2006MNRAS.370.2013S}. \cite{2006MNRAS.370.2013S} gave a list of possible candidates by searching for eclipsing binaries positioned inside the \ds\, instability strip. Some new \ds\, stars in eclipsing binaries were discovered by the investigation of the candidate stars given in this list \citep[e.g. ][]{2007IBVS.5798....1S, 2008CoAst.157..379S, 2011NewA...16...72S}. The current number of the eclipsing binary \ds\, stars is around 90 in the latest published catalog \citep{2017MNRAS.470..915K}. In recent years, this number is dramatically growing thanks to the high-precision data provided by the space telescopes such as Convection Rotation and Planetary Transits (CoRot, \citeauthor{2006cosp...36.3749B} \citeyear{2006cosp...36.3749B}), \textit{Kepler} \citep{2010Sci...327..977B} and Transiting Exoplanet Survey Satellite (TESS, \citeauthor{2014SPIE.9143E..20R} \citeyear{2014SPIE.9143E..20R}). In addition to revealing the new eclipsing binary \ds\, stars, space telescopes also uncovered some new properties of these systems. For example, the first observational proof of \ds\, type pulsations in only one hemisphere of the pulsating component of a close binary was discovered with TESS data \citep{2020NatAs...4..684H}. Afterwards, a similar object was also found in an eclipsing binary \citep{2021MNRAS.503..254R}. It was shown that in these close binaries the oscillation axis of the pulsating binary component is aligned with the tidal axis and hence the star exhibits pulsations more effectively on one of its hemisphere \citep{2020MNRAS.494.5118K, 2021MNRAS.503..254R}. These variables are now dubbed ``single-sided pulsators'' \citep{2020NatAs...4..684H} or, more general, tidally tilted pulsators \citep{2020MNRAS.498.5730F}. As TESS observed almost the entire sky \citep{2015JATIS...1a4003R}, it has supplied sufficient data which help us to deeply understand the nature of stars like single-sided pulsators. TESS also observed a lot of known eclipsing binaries with \ds\, stars which had only ground-based photometric data of much lower precision until now. One of these systems is TZ\,Dra. 

TZ\,Dra is an eclipsing binary system with a spectral type of A7\,V \citep{1960ApJ...131..632H, 1986PASP...98..690F}. Its pulsational characteristics were first revealed by \cite{2005ASPC..333..197M}. They found that the system exhibits a $\sim$28 min oscillation and classified the primary (hot) binary component as a \ds\, star. Pulsating stars in eclipsing Algols are called ``oscillating Eclipsing Algol (oEA)'' \citep{2004A&A...419.1015M}. TZ\,Dra is also an oEA system with an orbital period of 0.8660307\,d \citep{2004AcA....54..207K}. 
The oEAs are substantial systems to trace the effects of mass transfer and/or mass loss to pulsations. In these Algol-type systems, mass is transferred from the evolved cool component to the pulsating hot component \citep{1955AnAp...18..379K}. It was shown that mass transfer affects the pulsations \citep{2004A&A...419.1015M, 2004MNRAS.347.1317R}. The effect of mass transfer in Algols can be revealed by following different features of the stellar spectra. Mass transfer can show itself as absorptions and/or emissions in H$\alpha$ lines \citep{2001AJ....121.2723V, 2007MNRAS.379.1533S}. To investigate the effect/impact of mass transfer and/or mass loss on pulsations in oEA systems, TZ\,Dra is a valuable system. Therefore, we performed a detailed analysis of TZ\,Dra using its high-quality TESS data in combination with newly acquired high-resolution spectra.

We organise the study as follows. In Sect.2, the information about the observational data is given. The radial velocity measurements and their analysis are introduced in Sect.3. In Sect.4 and 5, we present the spectral analysis and the binary modelling, respectively. The examination of the structure of the system's H$\alpha$ line is given in Sect.\,6. The orbital period changes and the time-series analysis are introduced in Sect.7 and 8, respectively. In the final section, Sect.9, discussion and conclusions are presented.

\section{Observational Data}

In our investigation, we used TESS data and obtained high-resolution spectra for TZ\,Dra. 

TESS originally was designed to identify new exoplanets in bright stars (I$_{c}$\,$<$\,13). It was launched in April 2018 and until now TESS has observed almost entire of the sky. The TESS data have been taken with 2-min short cadence (SC) and 30-min long cadence (LC) \citep{2015JATIS...1a4003R} during the nominal mission (first two years). Observations were done in 26 partly overlapping sectors of $24^{\circ} \times 96^{\circ}$ that were observed for about 27 days each. Since the start of the extended mission in July 2020, observations were done in a similar way but the cadence of the LC data has been increased to one observation every 10 minutes and a new 20-sec cadence has been introduced. The TESS data are released in the Barbara A. Mikulski Archive for Telescopes (MAST)\footnote{https://mast.stsci.edu}. For TZ\,Dra, only the LC data of sector 14 and SC data of sectors 25, 26 are available in the MAST. Simple Aperture Photometry (SAP) flux of both high level science product (HLSP) LC and SC TESS data were taken from the MAST and used in the study.

The spectroscopic data were obtained using the High Efficiency and Resolution Mercator \'{E}chelle spectrograph (HERMES). The HERMES is attached to the 1.2-m Mercator telescope at the Roque de los Muchachos observatory at La Palma, Spain. The spectrograph has a resolving power of 85000 and it provides spectra in a spectral range from 377 to 900\,nm \citep{2011A&A...526A..69R}. For TZ\,Dra, we gathered 36 HERMES spectra between May and September 2020. The spectra were taken at different epochs to get spectra spread over the orbital period of TZ\,Dra. The signal-to-noise (S/N) ratio of these spectra ranges from 60 to 100 and has an average of 83. 

In this study, the TESS photometric data and HERMES spectra were used for the radial velocity analysis, binary modelling, the determination of the atmospheric parameters such as effective temperature (\teff), surface gravity (\logg), microturbulence velocity ($\xi$) and also projected rotational velocity (\vsini) of the system and time-series analysis.

\section{Radial Velocity analysis}

To accurately determine the fundamental stellar parameters of system, binary modelling should be done with the radial velocity (\vr) measurements in addition to photometric data. Therefore, before performing the binary modelling we obtained the \vr\, data of the binary components. In this study, to measure the \vr\, values we used a synthetic spectrum that was generated considering the spectral type and an estimated \vsini\, value (~80\,\kms) of TZ\,Dra. The initial \vsini\, was determined by profile fitting. By using a synthetic spectrum, which has a similar spectral type and \vsini, possible errors in the radial velocity measurements were decreased. The \vr\, values were obtained utilizing the FXCOR task of the IRAF\footnote{http://iraf.noao.edu/} program package \citep{1986SPIE..627..733T}. This task uses the cross-correlation technique which compares one spectrum to a template spectrum with a known velocity and zero redshift to find the redshift and velocity dispersion of the system under investigation to measure \vr\, values. As a result of our analysis, as we did not find a secondary peak in the cross-correlation profiles, we could only measure the \vr\, values of the primary component. The resulting \vr\, measurements are listed in Table\,\ref{vr_table}.

\begin{figure}
\centering
\includegraphics[width=8.5cm, angle=0]{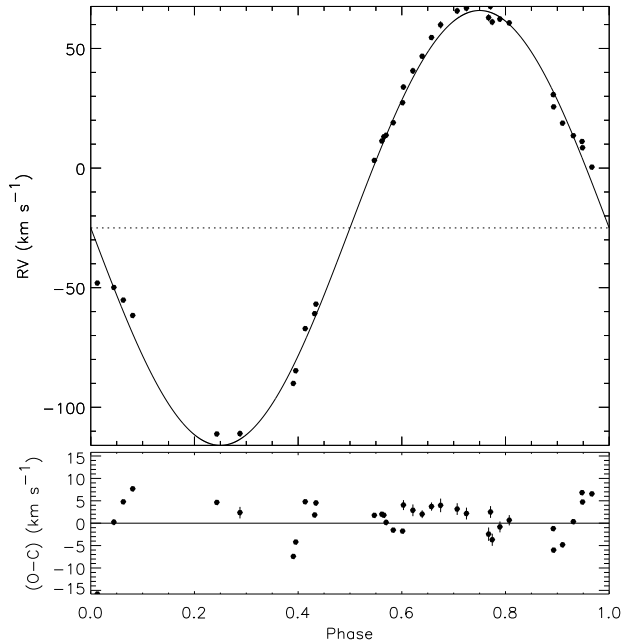}
\caption{Upper panel: Theoretical (solid line) fit to the \vr\, measurements (dots). Lower panel: residuals.}
\label{rvfit}
\end{figure}

The \vr\, analysis was carried out using the \texttt{rvfit} code\footnote{http://www.cefca.es/people/riglesias/rvfit.html} which is suitable for the \vr\, analysis of single and double-lined binary systems. First, an initial analysis was performed by fixing the orbital period of the system as 0.8660307\,$\pm$\,0.0000010 days \citep{2004AcA....54..207K} and the periastron passage time (T$_{o}$), the velocity of mass centre (V$_{\gamma}$), the amplitude of the \vr\, curve ($K$), argument of periastron ($\omega$), orbital eccentricity ($e$) and the projected semi-major axis (asin\textit{i}) parameters were searched for. As the $e$ value was found to be 0 in the first a few iterations, this value and also the $\omega$ parameter were assumed to be 0 and 90 degree respectively in the further analysis. After we approached the final values of the parameters the orbital period was investigated as well. In conclusion, we obtained the orbital parameters of TZ\,Dra that are given in Table\,\ref{rv_result}. The best theoretical \vr\, fit to the \vr\, measurements is shown in Fig.\ref{rvfit}.

\begin{table}
\centering
\caption{The results of the radial velocity analysis. $^{a}$ represents the fixed parameters.}
  \label{rv_result}
\begin{tabular*}{0.75\linewidth}{@{\extracolsep{\fill}}lc}
\hline
  Parameters  & Value  \\
\hline 
$P$ (d)        & 0.8660310$^{a}$\\
T$_{0}$ (HJD)  & 2452500.63606 $\pm$ 0.00004\\
$e$            & 0.0$^{a}$ \\
$\omega$ (deg)		&90$^{a}$ \\
$\gamma$ (km/s)	&-25.02 $\pm$ 0.11\\
$K_1$ (km/s)		&90.90 $\pm$ 0.20\\
$a_1\sin i$ ($R_\odot$)	&1.555 $\pm$ 0.003\\
$f(m_1,m_2)$ ($M_\odot$)	&0.0674 $\pm$ 0.0004\\
\hline
\end{tabular*}
\label{abunresult}
\end{table}

\section{Spectral analysis}

In our radial-velocity analysis, no evidence of the cool (secondary) component spectrum was found in the cross-correlation function. Therefore, in our spectral analysis, we determined only the atmospheric parameters of the hot (primary) component. In any case, we used the spectra taken at close to 0.5 orbital phase where the secondary component is partially covered by the primary. In this way, we decreased the possible effect of the secondary component in the spectra. Three spectra at around $\sim$0.5 phase (see, Table,\ref{vr_table}) were chosen and combined to achieve a higher S/N ratio ($\sim$165) spectrum.

Using the combined spectrum, we first determined the initial \teff\, value from the hydrogen line modelling. In this and next spectral analysis, the plane-parallel, hydrostatic, local thermodynamic equilibrium (LTE) ATLAS9 atmosphere models \citep{1993KurCD..13.....K} were used and the theoretical spectra were generated by using the SYNTHE code \citep{1981SAOSR.391.....K}. In the hydrogen line analysis, we assumed solar abundance and fixed the \logg\, as 4.0\,dex. These assumptions were made because the profile of the hydrogen lines is not affected by the metallicity and for the stars cooler than 8000\,K there is a negligibly weak dependence on \logg\, \citep{2002A&A...395..601S, 2005MSAIS...8..130S}. Additionally, in the hydrogen line analysis, we only used H$\delta$ and H$\gamma$ lines because it is known that in Algol type systems due to the mass transfer effect there could be additional absorption and/or emission in the H$\alpha$ and H$\beta$ lines \citep{2001AJ....121.2723V, 2007MNRAS.379.1533S}. Those extra features in the hydrogen lines can be analysed after subtracting the theoretical hydrogen line profile. This examination is presented in Sect.\,6. 

\begin{figure}
\centering
\includegraphics[width=8cm, angle=0]{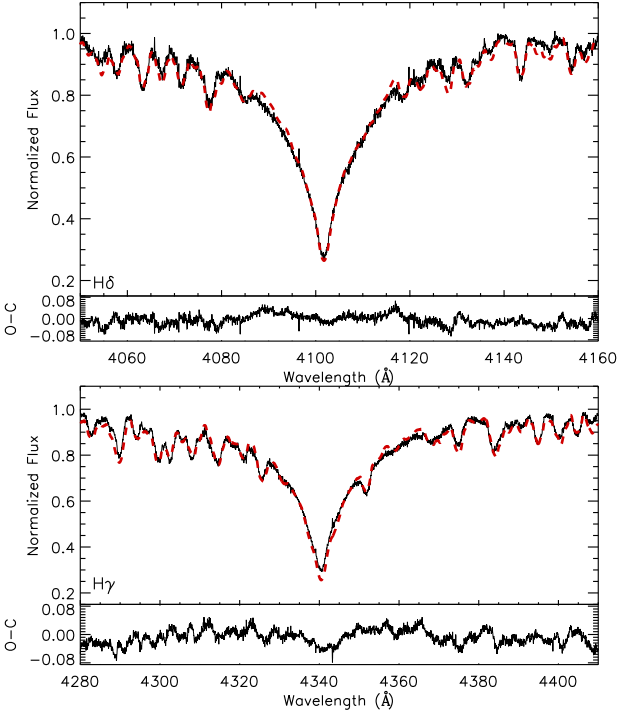}
\caption{Theoretical hydrogen line fit (dashed line) to the observed H$\delta$ and H$\gamma$ lines. The differences between the observed (O) and calculated (C) spectra are shown below of each panels.}
\label{hlinefit}
\end{figure}

In the hydrogen line analysis, several synthetic spectra were generated for a range of \teff\, ($\sim$ $7000-8500$\,K) in 100\,K steps and they were fitted to the observed hydrogen lines. The best fit theoretical spectra were found considering the minimum $\chi^2$ value (for more detail see \cite{2004A&A...425..641C}). As a result of the analysis, we found a hydrogen \teff\, values of 7700 $\pm$ 250\,K. The consistency of the theoretical hydrogen lines with the observations is demonstrated in Fig.\,\ref{hlinefit}.

To derive the \logg\, and $\xi$ values and improve the \teff\, value, we used the excitation/ionization-abundance relations which rely on the Boltzman and Saha equations. Taking into account the hydrogen-based \teff\, we decided to use iron (Fe) lines because of their significant number at this \teff. The method used in \cite{2016MNRAS.458.2307K} was applied in the same way in this study. First, the Fe lines were chosen from the line list of Kurucz\footnote{kurucz.harvard.edu/linelists.html} and their corresponding abundances were derived for a $T_{\rm eff}$, $\log g$, and $\xi$ range with a step of 100\,K, 0.1\,dex, and 0.1 \kms, respectively. The spectrum synthesis method was used in the analysis. It is known that for the correct atmospheric parameters, for different excitation/ionization potential one should get the same abundances as described by \cite{2016MNRAS.458.2307K}. In this way, the atmospheric parameters of the hot (primary) component of TZ\,Dra were obtained. The derived parameters are given in Table\,\ref{atmpar_result}. 

\begin{table}
\centering
\caption{The final atmospheric parameters and $v \sin i$ value of the hot (primary) component star of TZ\,Dra.}
\label{atmpar_result}
\begin{tabular*}{0.9\linewidth}{@{\extracolsep{\fill}}lccc}
\hline
$T_{\rm eff}$\,(K)     & $\log g$\,(cgs)     & $\xi$\,(km\,s$^{-1}$)   & $v \sin i$\,(km\,s$^{-1}$) \\
\hline
8100 $\pm$ 200 & 4.1 $\pm$ 0.2 & 1.9 $\pm$ 0.3 & 85 $\pm$ 3 \\
\hline 
\end{tabular*}
\label{abunresult}
\end{table}

\begin{table}
\begin{center}
\caption[]{Abundances of individual elements of the primary star and Sun (\citealt{2009ARA&A..47..481A}). Number of the analysed spectral lines is given in third column.}\label{abunresult}
 \begin{tabular}{lccc}
  \hline\noalign{\smallskip}
  Elements & Star abundance  & Number of &Solar abundance            \\
           &                 & spectral parts & \\
  \hline\noalign{\smallskip}
$_{6}$C   &8.63\,$\pm$\,0.45 & 1  & 8.43\,$\pm$\,0.05\\
$_{12}$Mg &7.98\,$\pm$\,0.40 & 4  & 7.60\,$\pm$\,0.04\\
$_{14}$Si &6.97\,$\pm$\,0.36 &5  & 7.51\,$\pm$\,0.03\\
$_{20}$Ca &6.48\,$\pm$\,0.31 &6 & 6.34\,$\pm$\,0.04\\
$_{21}$Sc &3.37\,$\pm$\,0.41 &3  & 3.15\,$\pm$\,0.04\\
$_{22}$Ti &4.91\,$\pm$\,0.34 &17 & 4.95\,$\pm$\,0.05\\
$_{23}$V  &4.15\,$\pm$\,0.42 &2  & 3.93\,$\pm$\,0.08			  \\
$_{24}$Cr &5.57\,$\pm$\,0.35 &10 & 5.64\,$\pm$\,0.04\\
$_{25}$Mn &5.02\,$\pm$\,0.33 &5  & 5.43\,$\pm$\,0.05			   \\
$_{26}$Fe &7.37\,$\pm$\,0.21 &40 & 7.50\,$\pm$\,0.04\\
$_{28}$Ni &6.18\,$\pm$\,0.38 &5 & 6.22\,$\pm$\,0.04\\
$_{38}$Sr &2.71\,$\pm$\,0.45 &1  & 2.87\,$\pm$\,0.07\\
$_{56}$Ba &2.39\,$\pm$\,0.42 &2  & 2.18\,$\pm$\,0.07\\
  \noalign{\smallskip}\hline
\end{tabular}
\end{center}  
\end{table}

\begin{figure}
\centering
\includegraphics[width=8cm, angle=0]{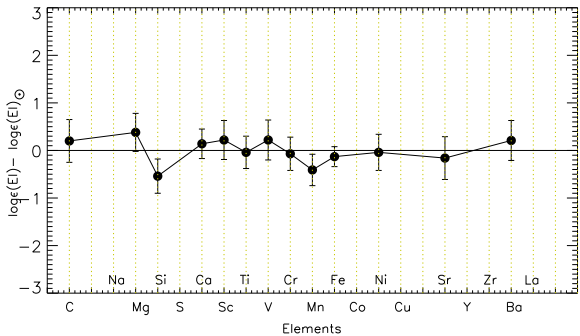}
\caption{The chemical abundance distribution of the primary component of TZ\,Dra. The solar abundance was taken from \citealt{2009ARA&A..47..481A}.}
\label{abun_dist}
\end{figure}

\begin{figure*}
\centering
\includegraphics[width=15cm,height=7cm, angle=0]{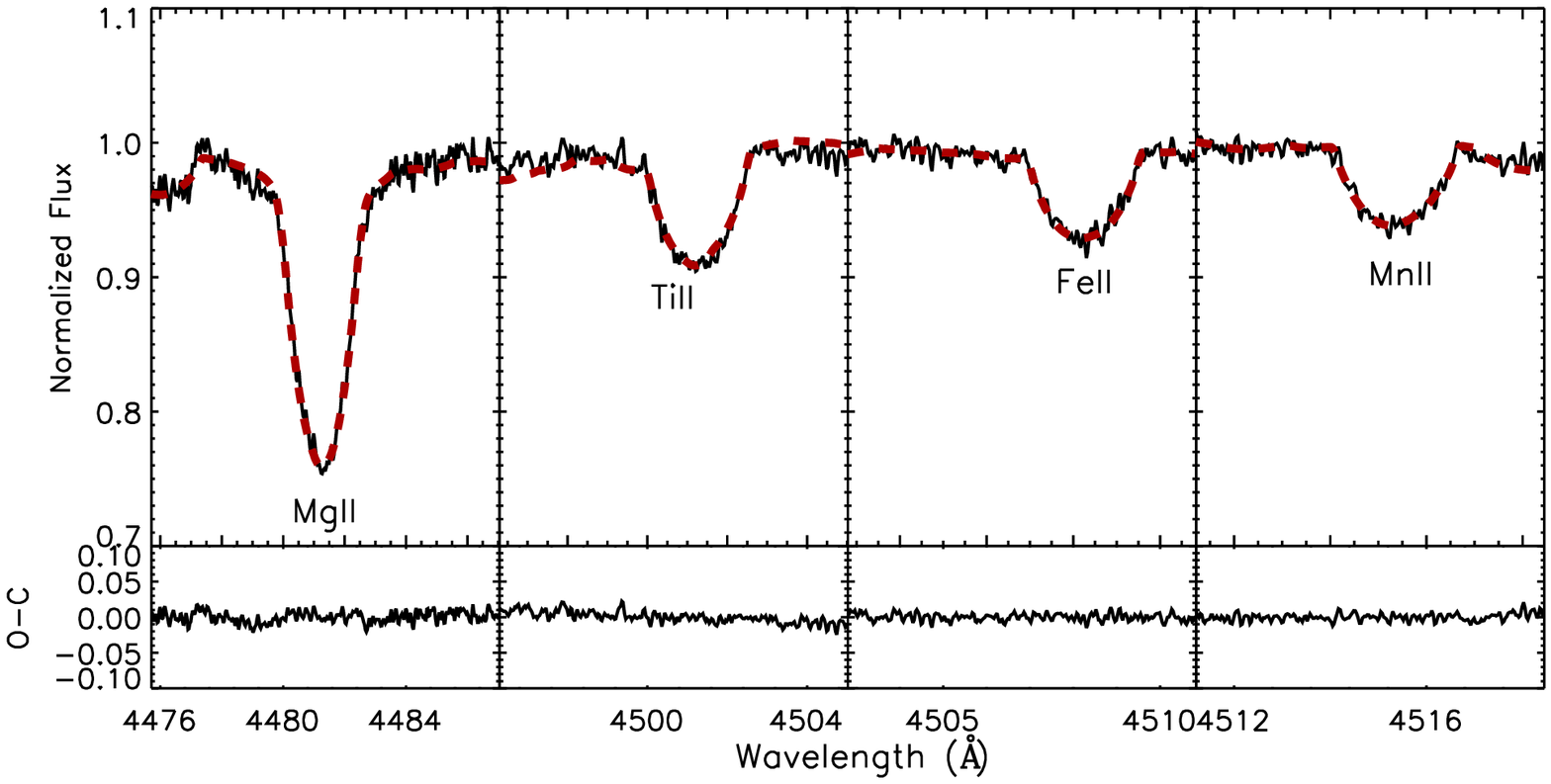}
\caption{Consistency between the synthetic (dashed lines) and the observed spectra.}
\label{abun_comp}
\end{figure*}

After we determined the atmospheric parameters, we derived the chemical abundances by taking these parameters as input and using the Kurucz line list.
In this analysis, the abundances of the individual elements were adjusted until the minimum in the difference between the calculated and observed spectra was found. During this analysis, the \vsini\, parameter was also searched for. The abundances of the chemical elements are listed in Table\,\ref{abunresult}, while the \vsini\, is given in Table\,\ref{atmpar_result}. The chemical abundances distribution is also illustrated in Fig.\,\ref{abun_dist}. The consistency between the resulting synthetic and the observed combined spectrum is shown in Fig.\ref{abun_comp}.

The errors in the atmospheric parameters were estimated considering the 5\% difference in the used relations in the analysis.  Additionally, the uncertainties in the chemical abundances were calculated taking into account the effects of the errors in the input parameters and S/N ratio as described by \cite{2016MNRAS.458.2307K}.

\section{Binary modelling}

In order to derive the orbital and the fundamental stellar parameters of TZ\,Dra, we modelled the binary light curve. First, all the available TESS data were taken into account to check whether it changes from one sector to another. Because of the possible existence of spots on the surface of component(s), the flux level of the binary light curve during non-eclipsing phases could differ from one sector to another. These differences never affect the binary parameters, but only change the position of the spot on the surface of component(s). In the data of TZ\,Dra, no significant variation was found between the level of the out of eclipse brightness in the two sectors. Therefore, in the binary modelling we used both sector 25 and 26 data. 

The pulsations in the system do not deeply affect the orbital/geometrical variation of TZ\,Dra. However, the dominant pulsations frequencies \footnote{First ten highest amplitude frequencies given in Table\,\ref{tab:puls_freq}} were cleaned from the light curve rather than multiples of orbital period by performing a preliminary Fourier analysis with {\sc Period04} \citep[][see Sect. 8.2 for a more detailed description]{2005CoAst.146...53L}. The remaining data were first phased, normalised and then binned to make them usable for the binary modelling with the Wilson-Devinney code \citep{1971ApJ...166..605W}. 

In this analysis, we used the findings of the previous binary modelling \citep{2013Ap&SS.343..123L} as input. Only the \teff\, value of the primary component and the results of the \vr\, analysis ($\gamma$, $e$, $a$) were taken from our study and they were fixed during the analysis. The \teff\, value of the primary star was found slightly different in two different approaches. Therefore, in this analysis, we used the average \teff\, (7900\,K) as input. The other fixed parameters are bolometric albedos \citep{1969AcA....19..245R}, bolometric gravity-darkening coefficient \citep{1924MNRAS..84..665V}, and logarithmic limb darkening coefficient \citep{1993AJ....106.2096V}. The TESS passband is not incorporated into the Wilson-Devinney code, therefore, the Cousins $I_c$-band was assumed in the binary modelling as the TESS passband is centered at Cousins $I_c$-band. The fixed coefficients were also chosen considering the $I_c$-band. Those coefficients were taken with the the same way as given \cite{2019MNRAS.488.5279K}. The \teff\, of the cool component, phase shift ($\phi$), orbital inclination ($i$), possible third body light contribution ($l$$_{3}$), the mass ratio ($q$=$M$$_{hot}$/$M$$_{cool}$), fractional luminosities and the dimensionless potential ($\Omega$) of binary components were used as free parameters. During the analysis, a semi-detached binary configuration was assumed as the TZ\,Dra system has been classified as such before \citep{2013Ap&SS.343..123L}. The Wilson-Devinney code combined with Monte-Carlo simulation was used in the binary modelling \citep{2004AcA....54..299Z, 2010MNRAS.408..464Z}. As the system is an Algol-type binary, there could be mass transfer between the components which was shown by \cite{2013Ap&SS.343..123L}. During mass transfers/losses, angular momentum changes and the rotational velocity of the components might change. Therefore, we carried out the modelling by testing non-synchronous and synchronous rotation. As we obtained the better $\chi^2$ value for the synchronous rotation we worked with this assumption throughout the binary modelling.

\begin{figure}
\centering
\includegraphics[width=8cm, angle=0]{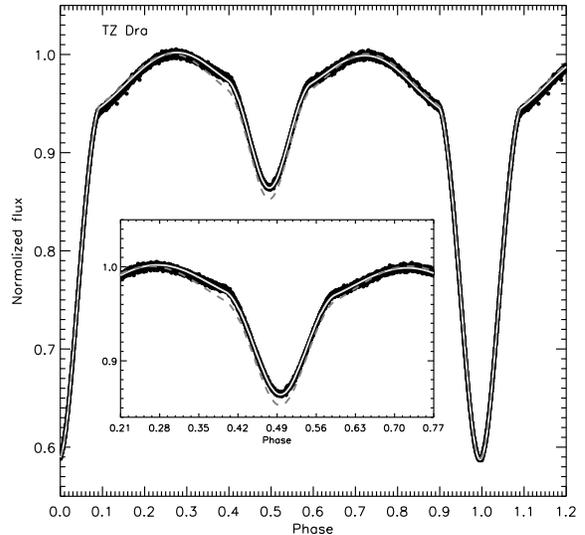}
\caption{Theoretical binary modelling fit with spot assumption (solid line) and without spot assumption (dashed-line).}
\label{lc_fit}
\end{figure}

\begin{figure}
\centering
\includegraphics[width=8cm,height=6cm, angle=0]{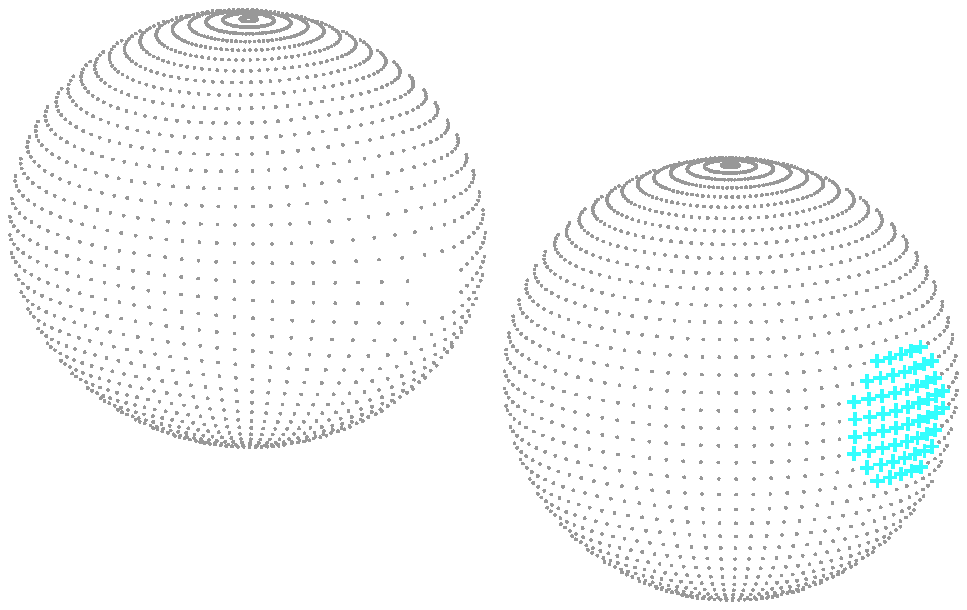}
\caption{Roche geometry of TZ\,Dra at around 0.6 phase. Blue spot on the primary star represents the hot-spot.}
\label{roche}
\end{figure}

As the system was found to be a single-lined binary, there is no certain information about the $q$ value. Therefore, a $q$ search was carried out by checking the minimum $\chi^2$ value. Consequently, it was found that the $q$ value should be around 0.42\,$\pm$\,0.05. The q search is shown in Fig.\,\ref{qsearch}. The binary modelling was started adopting $q$ value input. In the first binary modelling steps we noticed that in the out of eclipse light curve there are some asymmetries which are probably caused by a star spot(s). Therefore, we included star-spot modelling in our analysis as well. As a result of this analysis, we found the best fitting model with a hot spot on the primary star. The results of the binary modelling are listed in Table\,\ref{lcresult} and the best fitting theoretical model to the TESS data is illustrated in Fig.\,\ref{lc_fit}. The Roche lobe geometry of the system with the hot spot on primary's surface is shown in Fig.\ref{roche}

\begin{table}
\begin{center}
\caption[]{Results of the light curve analysis and the fundamental stellar parameters. The Subscripts 1, 2 and 3 represent the primary, the secondary, and third binary components, respectively. $^a$ Shows the Fixed Parameters.}\label{lcresult}
 \begin{tabular}{lr}
  \hline\noalign{\smallskip}
   Parameter			  &  Value   	    \\
  \hline\noalign{\smallskip}
$i$ ($^{o}$)	       	          & 75.138 $\pm$ 0.015     \\	
$T$$_{1}$$^a$ (K)                 & 7900 $\pm$ 250  	   	\\	
$T$$_{2}$ (K)    	          & 4970 $\pm$ 233	    		\\
$\Omega$$_{1}$		          & 3.848 $\pm$ 0.046      	  \\
$\Omega$$_{2}$		          & 2.734 $\pm$ 0.035     	  \\
Phase shift             	  & -0.0044 $\pm$ 0.0001   	  \\
$q$                     	  & 0.428 $\pm$ 0.013        	  \\
$r$$_{1}$$^*$ (mean)              & 0.2953 $\pm$ 0.0036      \\
$r$$_{2}$$^*$ (mean)              & 0.3070 $\pm$ 0.0026      \\
$l$$_{1}$ / ($l$$_{1}$+$l$$_{2}$)  & 0.787 $\pm$ 0.016       	  \\
$l$$_{2}$ /($l$$_{1}$+$l$$_{2}$)  & 0.213 $\pm$ 0.016         \\
$l$$_{3}$                         & 0.0     			  \\
\multicolumn{2}{c}{Spot Parameters}\\

Co-Latitude (deg)                        & 90$^a$  \\
Longitude (deg)   & 190.84$\pm$ 0.14\\
Radius  (deg)          & 19.53$\pm$0.04\\
Temperature Factor$^{**}$  & 1.052$\pm$0.012\\
\multicolumn{2}{c}{Derived Quantities}\\
$M$$_{1}$ ($M_\odot$)	          & 2.039 $\pm$ 0.004  	    \\	
$M$$_{2}$ ($M_\odot$)	          & 0.872 $\pm$ 0.003       	\\
$R$$_{1}$ ($R_\odot$)	          & 1.611 $\pm$ 0.052      	  \\
$R$$_{2}$ ($R_\odot$)		  & 1.675 $\pm$ 0.038      	  \\
log ($L$$_{1}$/$L_\odot$)		  & 0.960 $\pm$ 0.033       		  \\
log ($L$$_{2}$/$L_\odot$)		  & 0.188 $\pm$ 0.054       		  \\
$\log g$$_{1}$ (cgs)              & 4.30 $\pm$ 0.02 	     			    \\
$\log g$$_{2}$ (cgs)               & 3.92 $\pm$ 0.03 	    			    \\
$M_{bol}$$_{1}$ (mag)      & 2.34 $\pm$ 0.07      		  \\
$M_{bol}$$_{2}$ (mag)	  & 4.27 $\pm$ 0.08       		  \\
$M_{V}$$_{1}$ (mag)	          & 2.32 $\pm$ 0.09      		  \\
$M_{V}$$_{2}$ (mag)	          & 4.50 $\pm$ 0.13       		  \\
$M_{TESS}$$_{1}$ (mag)	          & 2.35 $\pm$ 0.05      		  \\
$M_{TESS}$$_{2}$ (mag)	          & 3.77 $\pm$ 0.09       		  \\
  \noalign{\smallskip}\hline
\end{tabular}
\end{center}
 \begin{description}
     \centering
 \item[ ] * fractional radii, ** \teff$_{spot}$/\teff$_{star}$
 \end{description}
\end{table}

The fundamental stellar parameters such as $M$, $R$, luminosity ($L$), bolometric magnitude ($M_{bol}$), absolute magnitude ($M_{V}$) and also \logg\, parameters for both binary components were calculated using the mass function $f$ found in the \vr\, analysis, the well-known Kepler and Pogson equations. These parameters are also listed in Table\,\ref{lcresult}.

\section{H$\alpha$ line profile and mass transfer examination}\label{alp}

Algols are semi-detached, close, short-period binary systems that consist of a hot (B-A) main-sequence and a cooler, less massive evolved component. This evolved cooler component fills its Roche lobe and transfers mass from Roche lobe overflow \citep{1955AnAp...18..379K}. The mass transfer can be noticed from some additional features affecting the photometric and spectroscopic data. The effect of mass transfer in Algols shows itself especially as distortions in the \vr\, data of the primary component, and additional emission and/or absorption in the H$\alpha$ and H$\beta$ lines \citep{2001AJ....121.2723V, 2007MNRAS.379.1533S}. When the \vr\, curve, H$\alpha$ and H$\beta$ line profiles of TZ\,Dra were examined, we noticed a small distortion in the \vr\, data especially at around 0.75 phase and significantly different profile in H$\alpha$. As explained by \cite{2001AJ....121.2723V}, emissions are caused by stream-disk or star-stream interaction in a region between binary components. Absorption lines may also originate from the matter around the primary component or ``stream projected against primary'' \citep{2001AJ....121.2723V}. Therefore the existence of these absorption and emission profiles in H$\alpha$ gives us information about the mass transfer. Consequently, we extracted the H$\alpha$ line profiles of TZ\,Dra. 

\begin{figure*}
\centering
\includegraphics[width=15cm,height=10cm, angle=0]{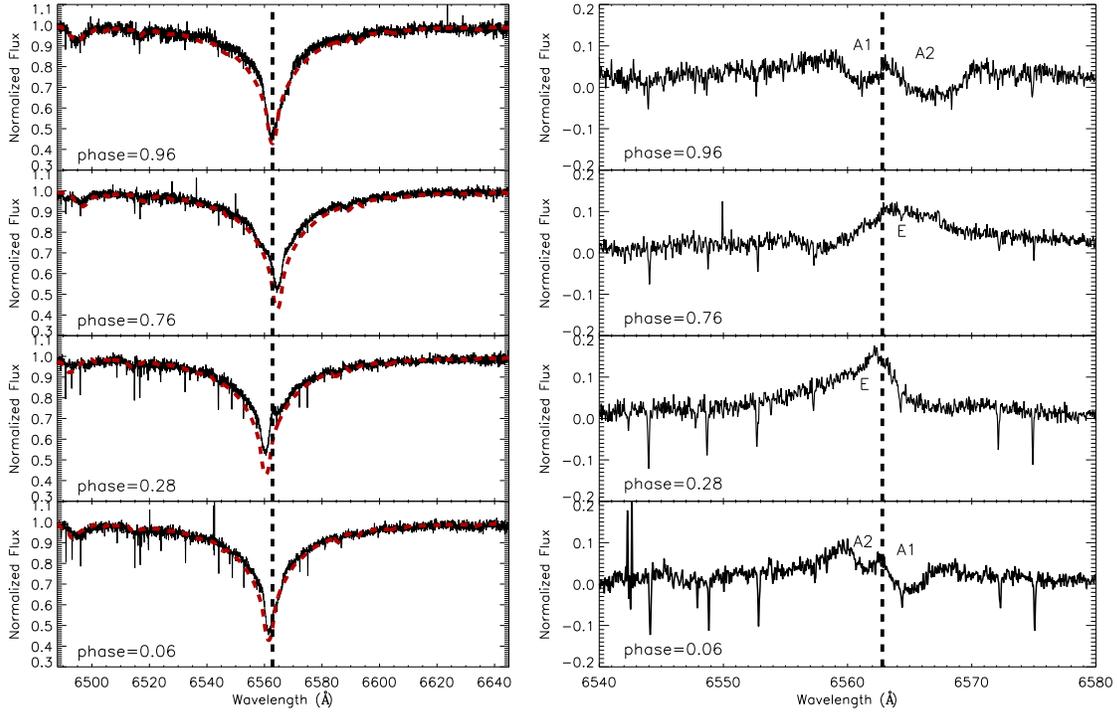}
\caption{The H$\alpha$ profiles at different orbital phases (left panels) and the residuals (right panels) from the observed and the theoretical H$\alpha$ (red dashed lines) profiles shown in the left panels. The vertical dashed lines represent the laboratory wavelength of H$\alpha$. The abbreviations A1, A2 and E show  absorption 1, absorption 2 and emission lines seen in the residuals.}
\label{halp_fig}
\end{figure*}

\begin{figure}
\centering
\includegraphics[width=8cm,height=6cm, angle=0]{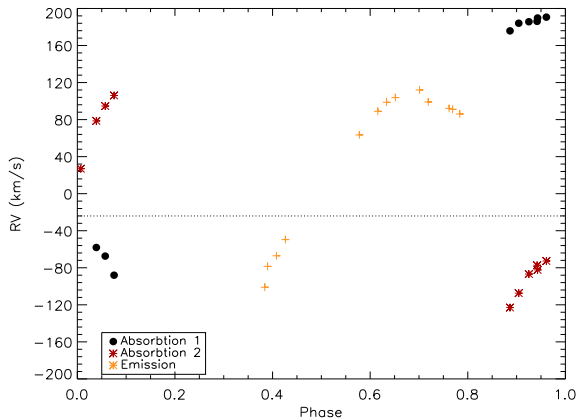}
\caption{The radial velocity variations of the absorption and emission lines shown in Fig.\,\ref{halp_fig}.}
\label{halp_rv}
\end{figure}

The best fit theoretical atmosphere model obtained in the analysis of the hydrogen lines was used in this analysis. This atmosphere model was fitted to the H$\alpha$ profiles and then it was extracted from the observed H$\alpha$ lines. In this way, the profiles of absorption and emission lines were obtained as shown in Fig.\ref{halp_fig}. In these residual spectra, we found one emission and two absorption line profiles. The \vr\, changes of these lines were measured and the variation of them according to the orbital phase were obtained as shown in Fig.\,\ref{halp_rv}. When the \vr\, variation of these lines were examined, we noticed that the  profile of emission is caused by the star-stream or star-disk interaction between two stars. These interactions create emission lines and they are visible at around near quadrature phases as obtained in this study \citep{2001AJ....121.2723V}. The absorption line 1 (see Fig.\ref{halp_fig}) is thought to be due to the stream projected against the primary component. In this case, absorption lines are observed near primary eclipse ingress or egress \citep{2001AJ....121.2723V} as we obtained for absorption line 1. Additionally, the \vr\, variations of this line follows the \vr\, profiles of the primary component. The other absorption line (line 2) seems to follow the secondary component and there is no explanation for this kind of absorption line profile occurring in Algol type systems. One interpretation could be that this line is caused by the mass stream from the secondary component. It only detected around the primary eclipse as can be seen from Fig.\ref{halp_rv}. As a result, with this investigation, we revealed the mass transfer signature of TZ\,Dra.

\section{Orbital period variation}
Mass transfer and mass loss in a system can be evidenced by an orbital period variation analysis as well. Hence, we also studied the orbital period changes. All available literature minima times of TZ\,Dra were collected from the O-C Gateway\footnote{http://var2.astro.cz/ocgate/}. Additionaly, we used TESS data to calculate new minima times. In this calculation, only the SC data were used because the sampling of these light curves allows precise determination of the times of minima. We measured three primary and secondary minima times from each sector and in total 12 new minima times were derived from the TESS observations. To obtain a new minima time a photometric observation was carried out at the \c{C}anakkale Onsekiz Mart Observatory with the IST60 telescope (60-cm). The minima times from the TESS and new photometric observation were calculated using the method of \cite{1956BAN....12..327K}. All the minima times are listed in Table\,\ref{minima_table}. To increase the sensitivity of our orbital-period variation (O-C) analysis, in addition to TESS and new minima times, we used only the literature CCD and photoelectric minima times in the analysis. In total, 16 photoelectric and 67 CCD minima times including the TESS minimas were used. During the analysis, data weights were taken as 3 for photoelectric data and 10 for CCD measurements and the weighted least squares method was used \citep{2009NewA...14..121Z}.

\begin{figure}
\centering
\includegraphics[width=8cm, angle=0]{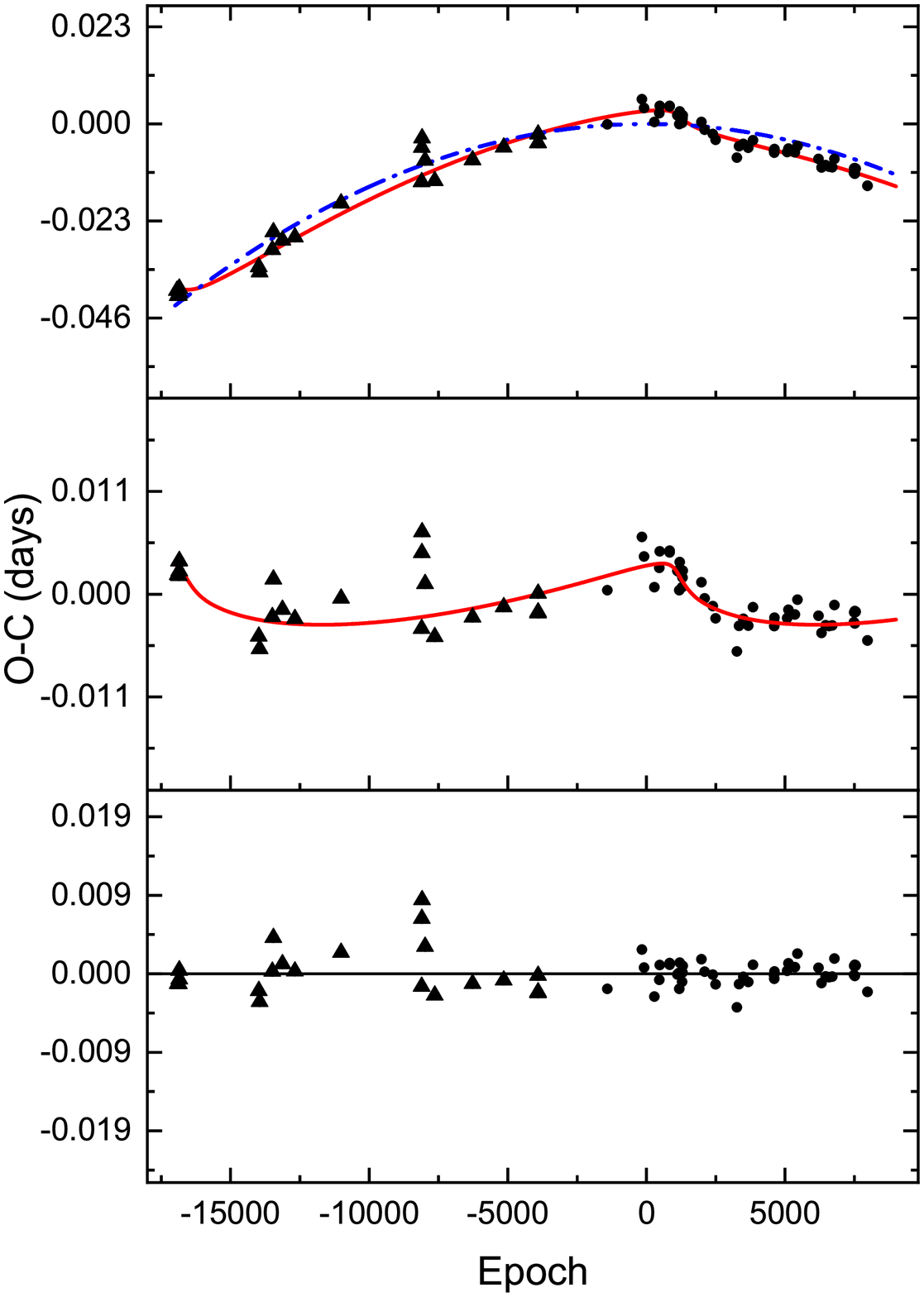}
\caption{The theoretical fits to the O-C data. The upper panel shows the both parabolic (blue dashed line) and the parabolic plus LITE model fit (red). The middle and lower panels illustrate the LITE fit alone and the residuals from the combined fit, respectively. The triangle and circle symbols represent the photoelectric and CCD data, respectively.}
\label{ocfig}
\end{figure}

 When we examined the O-C diagram of TZ\,Dra, we found a parabolic variation combined with another periodic change. First a parabolic fit was applied to the O-C data. As a result, the quadratic term ($Q$) was found to be -1.3265x10$^{-10}$ and a quadratic ephemeris was obtained as follows: 

\begin{math}
 HJD(MinI)\,=\,2452500.6419(5)\,+\,0.8660316(8)\,.\,E\,\
 -\,1.3265\,.\,10^{-10}(4)\,.\,E^{2}
 \end{math}\\


The parabolic variations in the O-C diagram could be caused by mass transfer between the binary component and/or mass loss from the system. For TZ\,Dra a 0.009\,s/year decrease in the orbital period was obtained as given in Table\,\ref{oc_table}. This shows us that the mass loss is effective in TZ\,Dra. We calculated the mass loss amount in the system considering the equation 16 of \cite{2014MNRAS.441.1166E}. This equation includes impacts such as dynamic and tidal effects. As a result, the mass loss amount was obtained as given in Table\,\ref{oc_table}. 

After removing the parabolic fit from the O-C data, we obtained another variation in the residuals. For the variation in the residual O-C diagram the light-time effect or light-travel time (LITE, \citeauthor{2009NewA...14..121Z}, \citeyear{2009NewA...14..121Z}) caused by a third body orbiting around the centre of mass \citep[e.g.][]{2005ASPC..335..103P, 2011NewA...16..498E} was tested. The physical relation between the O-C and LITE was formalised by \cite{1959AJ.....64..149I} who described the time delay ($\varDelta$T) in the period with the semi-major axis ($a$), speed of light (c), $i$, $e$, true anomaly ($\upsilon$) and longitude of periastron ($w$). The best theoretical fit representing this variation in the O-C diagram is obtained by the following equation:
\\

\begin{math}
 HJD(MinI)\,=\,2452500.6419(5)\,+\,0.8660316(8)E\,\ 
 -\,1.3265(4)\times10^{-10}E^{2} + \varDelta\,T 
\end{math}\\

 As a result, in the system we found a third component with a 45.8-year orbital period and a maximum mass of 0.17\,$\pm$\,0.03\,$M_\odot$ if it has an orbital inclination of 90$^{o}$. This third component could not found in the binary modeling because of it is relatively small mass. The results of the O-C analyses for the parabolic and LITE variations and the combined best theoretical fit to the observations are given in Table\,\ref{oc_table} and the best theoretical O-C fit to the observation points is shown in Fig.\,\ref{ocfig}. respectively. 

\begin{table}
\begin{center}
\caption[]{The results of the O-C diagram analysis. A and subscript 3 represent the amplitude of LITE and third body.}\label{oc_table}
 \begin{tabular}{lc}
  \hline\noalign{\smallskip}
  Parameter & Value          \\
  \hline\noalign{\smallskip}
$T$$_{o}$ (HJD) & 2452500.6419\,$\pm$\,0.0005\\
$P$ (day)       & 0.86603164\,$\pm$\,0.00000008\\
Q (day). (10$^{-10}$) & -1.3265\,$\pm$\,0.0003\\
dP/dT (s/year) & -0.009\,$\pm$\,0.001\\
dM/dt ($M_\odot$/year). 10$^{-9}$ & -3.52\,$\pm$\,0.24\\
P$_{3}$ (year) & 45.82\,$\pm$\,0.90\\
a$_{3}$sin$i$ (AU) & 0.98\,$\pm$\,0.09\\
A$_{s}$ (day) & 0.0032\,$\pm$\,0.0004\\
e$_{3}$ & 0.83\,$\pm$\,0.05\\
$\omega$$_{3}$ (degree) & 170\,$\pm$\,7\\
$f$ (m$_{3}$) ($M_\odot$) & 0.0005\,$\pm$\,0.0001\\
$m$$_{3}$ ($M_\odot$) ($i$=90) & 0.17\,$\pm$\,0.03\\
  \noalign{\smallskip}\hline
\end{tabular}
\end{center}  
\end{table}

\section{Time-series analysis}

The effect of binary interaction on the pulsations of a pulsating mass-accreting Algol-type system, U\,Gru was discussed by \cite{2019ApJ...883L..26B}. They show how important such systems are to probe the binary interaction with tidal asteroseismology. The possible effect of binary interactions and mass-transfer would change the structure and evolution scenario of the pulsating component and these are missing in the stellar models. Therefore, such systems are valuable for the theoretical modelling of mass transfer effect and/or mass loss effect on pulsations. Hence we carried out a frequency analysis of TZ\,Dra system. 

To perform a frequency analysis of the stellar pulsations we removed the mean binarity-induced light variations by a phenomenological fit consisting of the orbital frequency derived to be $1.1547060\pm0.0000004$\,d$^{-1}$ from the {\it TESS} photometry alone and its first 100 harmonics. 



\begin{figure}
\centering
\includegraphics[width=84mm,viewport=00 05 495 440]{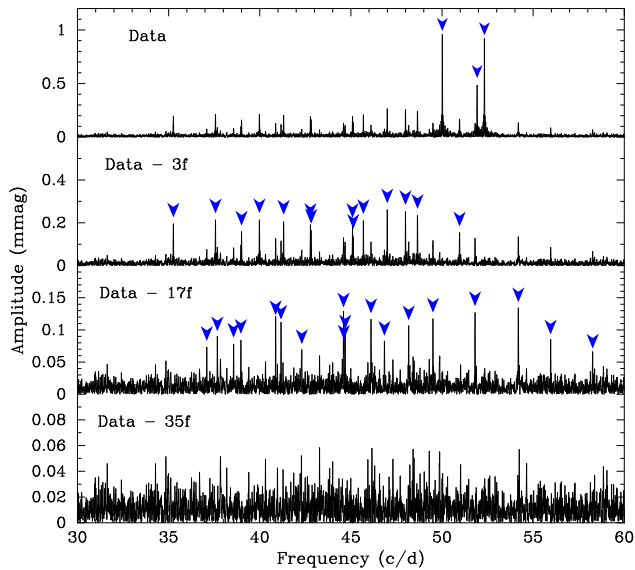}
\caption{The Fourier Transform of the residual {\it TESS} light curve of TZ Dra after filtering the binarity-induced light variations (top) and subsequent prewhitening steps. The blue arrows denote the signals detected.}
\label{fig:ft_plot}
\end{figure}

The residuals after the fit were then searched for frequencies of pulsation  using the {\sc Period04} software \citep{2005CoAst.146...53L}. This package applies single-frequency power spectrum analysis and simultaneous multi-frequency sine-wave fitting. It also includes advanced options such as the calculation of optimal light-curve fits for multiperiodic signals including harmonic, combination, and equally spaced frequencies. These sine-wave fits are subtracted from the data and the residuals examined for the presence of further periodicities. The application of this procedure, called prewhitening, to TZ Dra is illustrated in Fig.\,\ref{fig:ft_plot}.

To decide whether or not a periodic signal in a time series is statistically significant or not, several criteria have been proposed, the most commonly used probably the one by \citet{1993A&A...271..482B}. This criterion states that a given peak must exceed the mean amplitude in the Fourier spectrum by a factor of 4 in the local frequency domain to be considered significant. For space-based data, this can however lead to an overinterpretation of the periodic content \citep{2014MNRAS.439.3453B}. Consequently, we have stopped the frequency search after the detection of 35 pulsational signals to err on the side of caution; the strongest residual peak in the lowest panels of Fig.\,\ref{fig:ft_plot} has an amplitude S/N ratio of 4.4, with the noise calculated in a 10\,d$^{-1}$ window centred around it.

During the frequency analysis it became clear that many of the pulsational signals are spaced by integer multiples of the orbital frequency $\nu_{orb}$ from other pulsation frequencies. We therefore required that {\sc Period04} would fix those frequencies to exactly the predicted value as the orbital frequency is determined to considerably higher precision than the individual pulsation frequencies.
We list the frequency solution so derived in Table\,\ref{tab:puls_freq}, where we also quote formal errors on the derived parameters following \cite{1999DSSN...13...28M}.

\begin{table}
\centering
\caption{A least squares fit of the pulsation frequencies of TZ Dra. The zero point for the phases, $t_0 = 2459008.8585$, has been chosen to be at a time of primary eclipse. Error estimates for the independent frequencies and phases are given in braces in units of the last digits after the comma.}
\begin{tabular}{lccc}
\hline
&\multicolumn{1}{c}{frequency} & \multicolumn{1}{c}{amplitude} &
\multicolumn{1}{c}{phase}  \\
&\multicolumn{1}{c}{d$^{-1}$} & \multicolumn{1}{c}{mmag} &
\multicolumn{1}{c}{radians}   \\
& & \multicolumn{1}{c}{$\pm 0.012$} &
   \\
\hline
$\nu_1$ & 50.0174(1) & 0.950 &  4.51(1) \\
$\nu_1+2\nu_{orb}$ & 52.3268(1) & 0.910 &  1.35(1) \\
$\nu_2$ & 51.9264(2) & 0.467 &  0.83(2) \\
$\nu_3-2\nu_{orb}$ & 44.6863(12) & 0.092 &  1.84(11) \\
$\nu_3$ & 46.9957(4) & 0.264 &  4.95(4) \\
$\nu_4-\nu_{orb}$ & 45.6875(5) & 0.212 &  3.73(5) \\
$\nu_4$ & 46.8422(13) & 0.082 &  3.42(12) \\
$\nu_4+\nu_{orb}$ & 47.9969(4) & 0.254 &  3.73(4) \\
$\nu_5$ & 48.6545(5) & 0.235 &  4.91(4) \\
$\nu_5+2\nu_{orb}$ & 50.9639(7) & 0.154 &  1.88(7) \\
$\nu_6$ & 37.5715(5) & 0.218 &  3.16(5) \\
$\nu_6-2\nu_{orb}$ & 35.2621(6) & 0.195 &  3.06(5) \\
$\nu_7$ & 41.3112(5) & 0.207 &  4.71(5) \\
$\nu_7-2\nu_{orb}$ & 39.0018(7) & 0.149 &  4.86(7) \\
$\nu_8$ & 39.9832(5) & 0.214 &  6.26(5) \\
$\nu_8-2\nu_{orb}$ & 37.6738(12) & 0.089 &  3.00(11) \\
$\nu_9$ & 45.0894(6) & 0.194 &  1.75(5) \\
$\nu_9-2\nu_{orb}$ & 42.7800(6) & 0.190 &  4.88(5) \\
$\nu_{10}$ & 42.8204(7) & 0.167 &  2.88(6) \\
$\nu_{10}+2\nu_{orb}$ & 45.1298(8) & 0.139 &  6.15(7) \\
$\nu_{11}$ & 51.8128(9) & 0.128 &  4.10(8) \\
$\nu_{11}-2\nu_{orb}$ & 49.5034(9) & 0.118 &  4.21(9) \\
$\nu_{12}$ & 54.1882(8) & 0.134 &  4.86(8) \\
$\nu_{12}-7\nu_{orb}$ & 46.1053(9) & 0.117 &  1.66(9) \\
$\nu_{13}$ & 40.8723(10) & 0.114 &  4.03(9) \\
$\nu_{13}-2\nu_{orb}$ & 38.5629(14) & 0.079 &  4.25(13) \\
$\nu_{14}$ & 58.2668(18) & 0.062 &  2.87(16) \\
$\nu_{14}-2\nu_{orb}$ & 55.9574(13) & 0.084 &  5.81(12) \\
$\nu_{15}$ & 44.6175(14) & 0.076 &  3.90(13) \\
$\nu_{15}-2\nu_{orb}$ & 42.3081(16) & 0.069 &  4.14(15) \\
$\nu_{16}$ & 44.5955(9) & 0.126 &  2.23(8) \\
$\nu_{17}$ & 41.1696(10) & 0.107 &  2.44(10) \\
$\nu_{18}$ & 48.1700(10) & 0.106 &  4.19(10) \\
$\nu_{19}$ & 38.9667(13) & 0.085 &  2.53(12) \\
$\nu_{20}$ & 37.0940(15) & 0.074 &  1.21(14) \\
\hline
\end{tabular}
\label{tab:puls_freq}
\end{table}

The pulsation frequencies determined this way range from $35.26 - 58.27$ d$^{-1}$. From the mass and radius for the primary $\delta$ Scuti pulsator, one can derive pulsation ''constants'' $Q_i = P_i \sqrt{(\rho/\rho_{\odot})}$ \footnote{P$_{i}$: Period of the used frequency} between $0.012 - 0.020$\,d. This corresponds to pulsation in relatively high-order p modes in the range of the second to fifth radial overtone \citep{1981ApJ...249..218F} which, in combination with the relatively unevolved state of the pulsator, leads to the expectation of a pulsation spectrum with regular frequency structures.

Fourteen out of the 20 independent pulsation frequencies detected for TZ Dra occur in multiplets. Twelve of those are doublets spaced by twice the orbital frequency, one is a triplet with a weak centroid spaced by the orbital frequency. With one exception, all of the single frequencies are low in amplitude, so it is possible that they have undetected multiplet companions. Furthermore, many of the doublet components are fairly similar in amplitude, and all of them are, within the errors, either in phase or $\pi$\,rad out of phase at primary minimum.

\begin{figure*}
\centering
\includegraphics[angle=270,width=180mm,viewport=00 00 465 740]{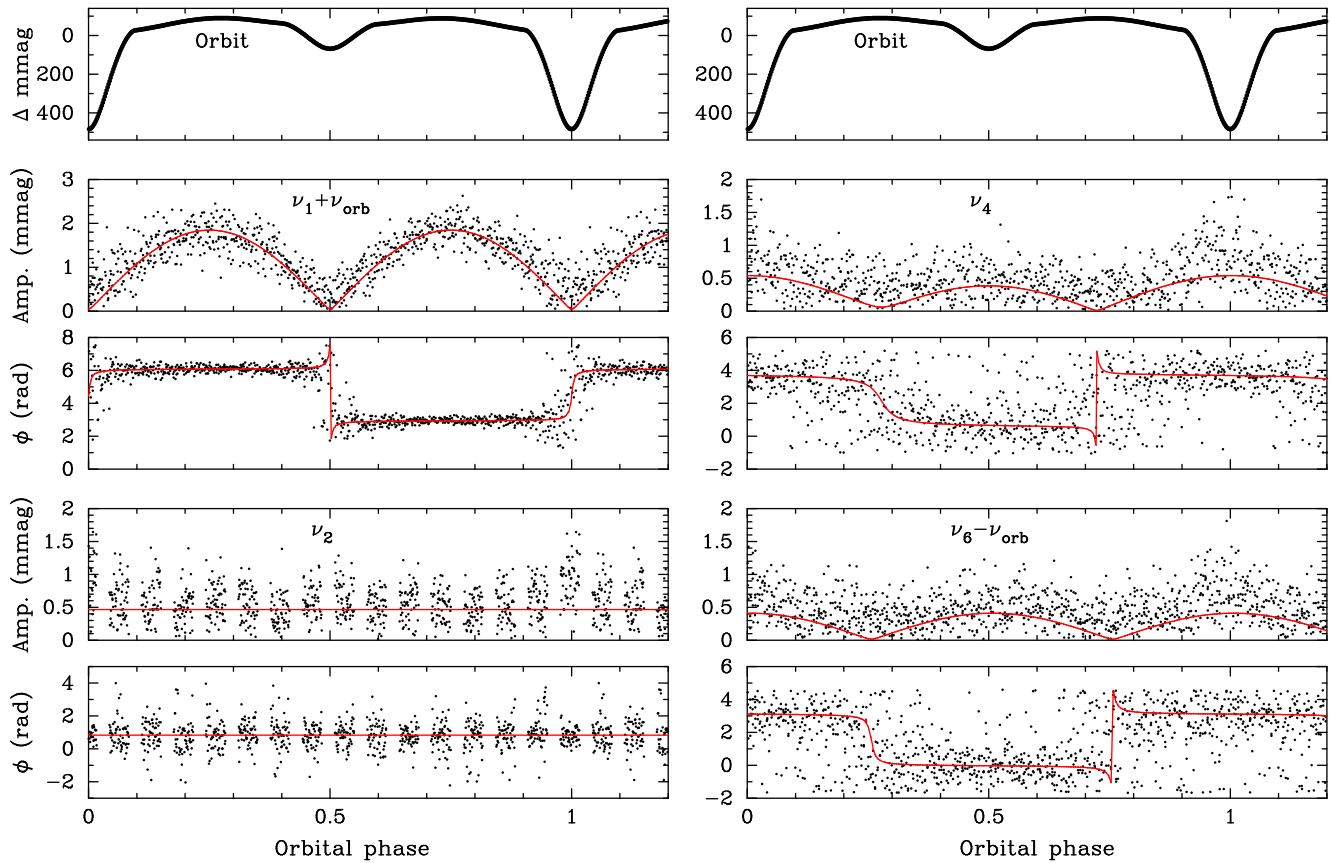}
\caption{The run of the pulsation amplitudes and phases of four selected modes of TZ Dra over the orbital cycle (bottom four panels left and right). The black dots are individual determinations, the red lines are fits using the signal parameters from Table\,\ref{tab:puls_freq}. Top: binary-induced light curve.}
\label{fig:ssp_plot}
\end{figure*}

These are signatures of tidally tilted pulsation. \cite{2005ApJ...634..602R} have shown how the oscillation spectrum of a single mode of given spherical degree $l$ and azimuthal order $m$ becomes modified due to the varying aspect of the pulsation axis over the orbital cycle. In brief, a single frequency splits up into multiplets determined by the type of mode, and by the inclination of the pulsational and rotational axes. These multiplets are spaced by integer multiples of the orbital frequency. This is what is seen here. The frequency doublets in our case thus are parts of multiplets centred on the frequency average of the two components. In Fig.\ref{fig:ssp_plot} we show the behaviour of the amplitudes and phases of a few selected modes over the orbit.

The amplitude of the $\nu_1$ doublet is at maximum near quadrature, and reaches almost zero near primary and secondary eclipse, respectively. The zero-crossing means that a node line of a nonradial pulsation mode crosses the line of sight. Assuming that the pulsational axis lies in the orbital plane (there would be no amplitude modulation if it was aligned with the rotational axis which we assume to be normal to the orbital plane), this must therefore be a sectoral mode. The singlet $\nu_2$ shows no phase modulation over the orbital cycle and almost no amplitude variation, except near primary eclipse. This would at first sight imply a radial mode. However, in that case one would expect the pulsation amplitude to decrease near primary eclipse, when part of the pulsating star is covered by the cooler companion that, at this orbital phase, contributes relatively more to the total flux observed. That this is not the case implies that $\nu_2$ is also due to a nonradial mode, and that the increase in amplitude during primary eclipse is due to part of the surface that causes geometrical cancellation \citep{1977AcA....27..203D} being covered by the companion. The mode triplet $\nu_4$ and the doublet including $\nu_6$ behave similarly: they have highest amplitude at primary and secondary eclipse, respectively, but an amplitude close to zero near quadrature. They are therefore tidally tilted axisymmetric modes.

Which of the mode doublets correspond to axisymmetric and sectoral modes, respectively, can be directly read off Table\,\ref{tab:puls_freq}. The multiplets that are in phase at primary minimum (triplet $\nu_4$ and the doublets containing $\nu_6$, $\nu_7$, $\nu_{11}$, $\nu_{13}$ and $\nu_{15}$) are the $m=0$ modes and the doublets $\pi$\,rad out of phase ($\nu_1$, $\nu_3$, $\nu_{5}$, $\nu_{8}$, $\nu_{9}$, $\nu_{10}$, $\nu_{12}$ and $\nu_{14}$) are sectoral modes. It is interesting to note that the centroid frequencies of the doublets containing axisymmetric modes are fairly equally spaced in frequency and may imply a mean frequency spacing of about 7.1\,d$^{-1}$ for consecutive radial orders of modes of the same $l$, which is close to the expected value of a star with the parameters of the primary of TZ Dra listed in Table\,\ref{lcresult}. Whereas a detailed seismic study of the pulsating component of TZ Dra is out of the scope of the present paper, and will need proper consideration of the tidal distortion of the oscillations (cf. \citealt{2020MNRAS.498.5730F}) and the altered stellar structure due to ongoing mass transfer \citep[e.g.,][]{2021MNRAS.505.3206M}, let it suffice to say that this is a very interesting system for more in-depth theoretical modelling.

In comparison to the remaining pulsational mode structure of TZ Dra, the frequency doublet $\nu_{12}$ and $\nu_{12}-7\nu_{orb}$ appears to be an ''outlier''. Perhaps the frequency match is just a numerical agreement, but in that case there would be little reason why the phases of those two signals are also $\pi$\,rad out of phase. This is a hint that these two signals are generated by the same pulsation mode, which then must however be $l>3$.

\begin{figure}
\centering
\includegraphics[width=8cm, angle=0]{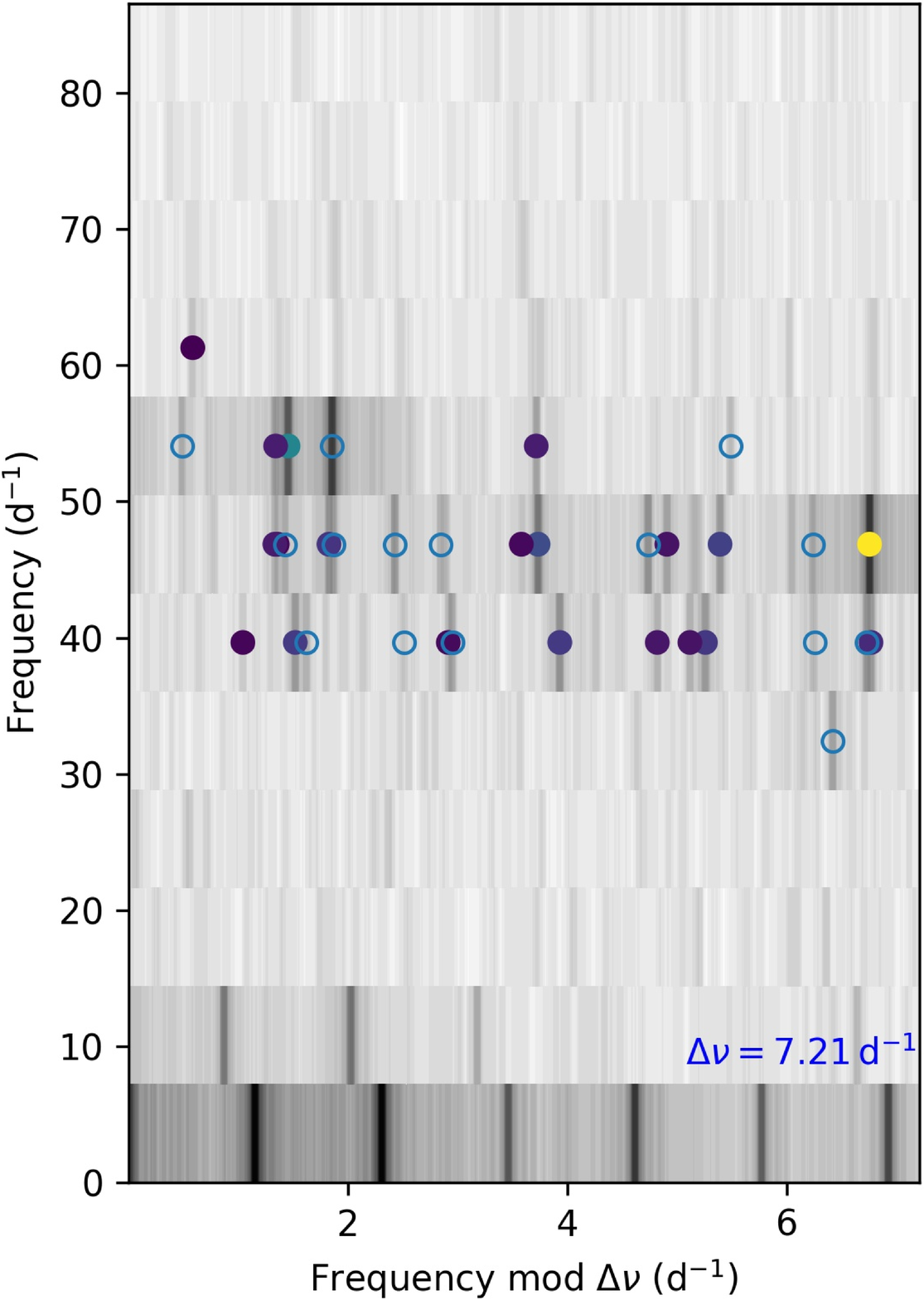}
\caption{The greyscale shows the Fourier amplitude spectrum of TZ\,Dra after removing the binary variations, shown in echelle format. The symbols show the peaks listed in Table\,\ref{tab:puls_freq}, colour-coded by amplitude. The open symbols are marked in Table\,\ref{tab:puls_freq} as being combinations with the orbital frequency. The figure was produced using the {\tt echelle} package \citep{2020zndo...3629933H}. }
\label{fig_echelle}
\end{figure}

The pattern of high-frequency pulsation modes in TZ Dra is similar to those found in young delta Scuti stars by \cite{2020Natur.581..147B}.  In Fig.\,\ref{fig_echelle} we show the Fourier amplitude spectrum of TZ Dra in \'{e}chelle format, using the residual light curve after removing the binary signal. Using a large separation of $\Delta\nu$=7.21\,d$^{-1}$, we see that the pulsation modes align in several vertical ridges, similar to the examples of more complicated echelle diagrams in Fig.\,4 of \cite{2020Natur.581..147B}.

\section{Discussion and Conclusions}

In this study, we present the results of the spectroscopic and TESS photometric examinations of TZ\,Dra. Using the high-resolution spectra, we measured the radial velocities of primary component and found no significant signal from the cool (secondary) component. Therefore, the spectral analysis was carried out only for the hot (primary) component. The TESS photometric data were used in the binary modelling and pulsational frequency analysis. 

The first binary modeling of the system was performed by \cite{2013Ap&SS.343..123L}. Since they used ground-based photometric data in their analysis, there are significant differences between the results of our binary modelling and theirs because of the sensitivity difference between the TESS and ground-based data. The $i$ (77.6$^{o}$) and $q$ (0.31) values show significant discrepancy, as a result the fundamental stellar parameters obtained in this study are quite different from the findings of \cite{2013Ap&SS.343..123L}. Additionally, in their analysis, no spot variation was found in the light curve. Thanks to the spectroscopic measurements and high-quality TESS data the fundamental stellar parameters for TZ\,Dra system were determined precisely. 


TZ\,Dra is an Algol-type system. In such semi-detached systems, it is known that the mass transfers from the evolved cool component to the hot primary component. The effects of this mass transfer can be seen in the photometric and spectroscopic data. Therefore, we traced the signature of mass transfer especially in the spectra of TZ\,Dra. It turned out that H$\alpha$ lines show quite different structures which allowed us to confirm the mass transfer. As a result of the H$\alpha$ line profile, we found that there is an emission line that is caused by the star-stream or star-disk interaction between both binary components. Additionally, it was obtained that there is a stream projected against the hot component. Furthermore, in the binary modelling, we also found a hot spot on the primary star's surface. 
  
 To explain the mass-loss in Algol-type systems there are four different mechanisms, bipolar jets \citep{2000A&A...358..229U, 2002A&A...391..609U}, mass-loss from the third Lagrangian point \citep{2007ARep...51..836S}, winds \citep{2009PhT....62i..52M} and the presence of a hot spot \citep{2013A&A...557A..40D, 2015A&A...577A..55D}. The hot spot mechanism seems to be the more suitable and convincing mechanism to explain mass loss from Algols. According to the hotspot model, the mass should be ejected from the primary star's surface via the radiation pressure of the hotspot \citep{2011A&A...528A..16V, 2013A&A...557A..40D, 2015A&A...577A..55D}. To reveal possible mass loss from the system an orbital period variation analysis was performed as well. In conclusion, a decrease in the orbital period was obtained and that could be explained with the mass loss via the hotspot and the winds in the secondary component. With these results, we showed that TZ\,Dra is a important system to understand the mass transfer and the mass loss mechanism in Algol-type binaries.
 
The pulsation feature of the primary binary component was examined and high pulsation frequencies (i.e. $p$ modes) ranging from 35.26 to 58.27 d$^{-1}$ were found. During the analysis, it was found that significant amount of pulsation frequencies are spaced by integer multiples of the orbital frequency and the star was shown to be a tidally tilted pulsator. According to the pulsation analysis we estimated that the star show high-order p modes in the range of second to fifth radial overtones. Good evidence for a regular pulsational frequency spacing of about 7.2\,d$^{-1}$ was found both from the tidally tilted pulsations and an \`{e}chelle diagram. 
 
TZ\,Dra is an interesting object for understanding the mass transfer and mass loss mechanism in Algol-type systems, as well as its impact on the pulsations. With the combination of the results of the studies for similar systems, these effects can be interpreted more efficiently and help us improve the current knowledge about the Algol type pulsating systems.  

\section*{Acknowledgments}
The authors would like to thank the reviewer for useful comments and suggestions that helped to improve the publication. This study  has  been  supported by  the  Scientific  and  Technological  Research  Council (TUBITAK) project through 120F330. GH thanks the Polish National Center for Science (NCN) for supporting the study through grant 2015/18/A/ST9/00578. IST60 telescope and its equipment are supported by the Republic of Turkey Office of Strategy and Budget with the project 2016K12137 and Istanbul University with the project numbers BAP-3685 and FBG-2017-23943.  TRB acknowledges support from the Australian Research Council through Discovery Project DP210103119. Based  on  observations  made  with  the Mercator  Telescope,  operated  on  the  island  of  La  Palma  by  the Flemish Community, at the Spanish Observatorio del Roque de los Muchachos of the Instituto de Astrof\`{\i}sica de Canarias. The TESS data presented in this paper were obtained from the Mikulski Archive for Space
Telescopes (MAST). Funding for the TESS mission is provided by the NASA Explorer Program. This work has made use of data from the European Space Agency (ESA) mission Gaia (http://www.cosmos.esa.int/gaia), processed by the Gaia Data Processing and Analysis Consortium (DPAC, http://www.cosmos.esa.int/web/gaia/dpac/consortium). Funding for the DPAC has been provided by national institutions, in particular the institutions participating in the Gaia Multilateral Agreement. This research has made use of the SIMBAD data base, operated at CDS, Strasbourq, France. 

\section*{DATA AVAILABILITY}
The data underlying this work will be shared at reasonable requestto the corresponding author.

 \appendix
 \begin{figure}
\centering
\includegraphics[width=8cm,height=6cm, angle=0]{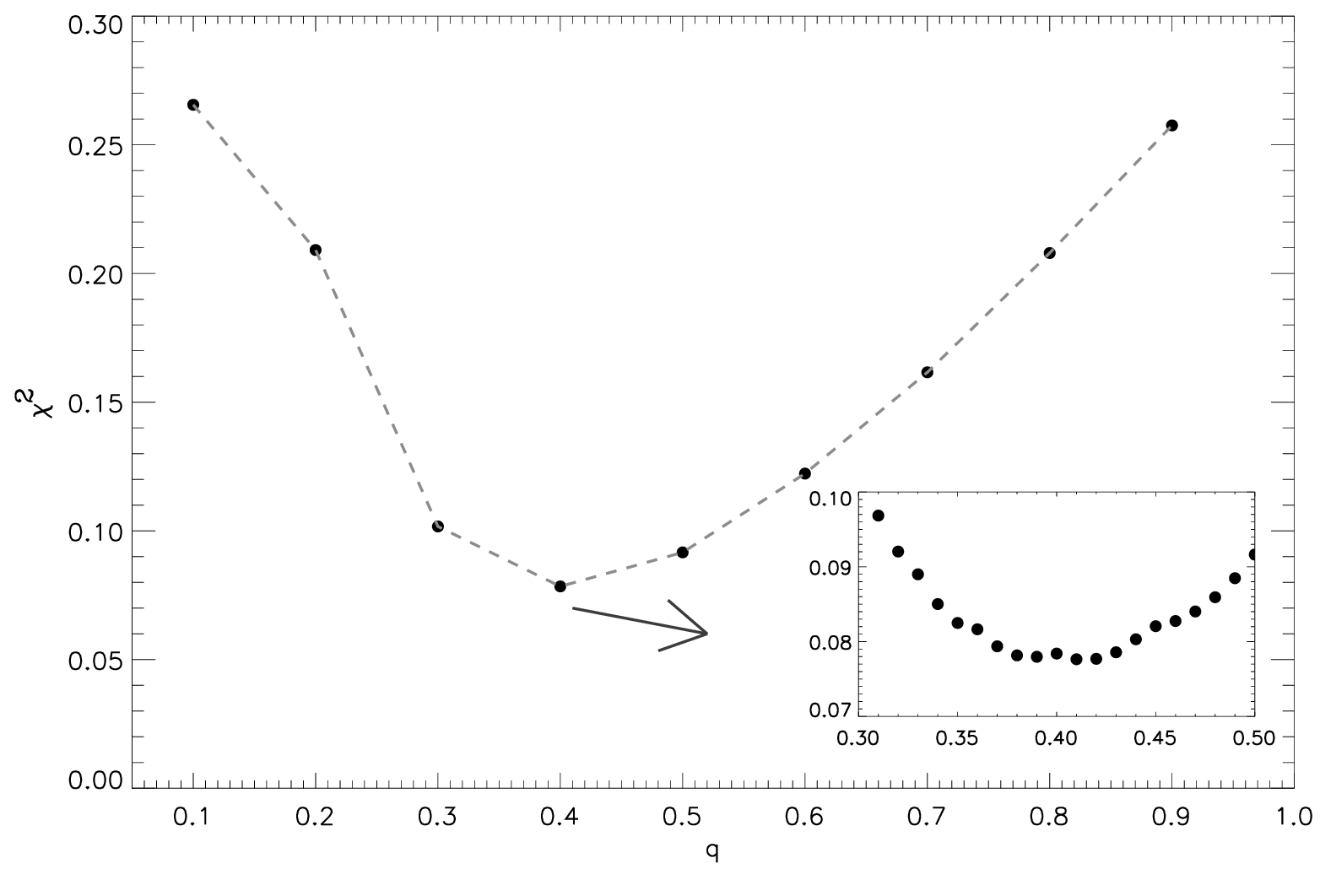}
\caption{q search for TZ\,Dra.}
\label{qsearch}
\end{figure}

\begin{table}
\begin{center}
\caption[]{Values of the \vr\, measurements. The $^{*}$ symbol show the spectra used in the spectral analysis according to the orbital phase.}\label{vr_table}
 \begin{tabular}{lcc}
  \hline\noalign{\smallskip}
  HJD        & phase &\vr\, \\
  (2450000+) &       &(\kms) \\
  \hline\noalign{\smallskip}
8986.67841 & 0.39 & -90.041 $\pm$ 0.327 \\
8988.55955 & 0.56$^{*}$ &-11.297 $\pm$ 0.331 \\
8990.68251 & 0.01 &-48.080 $\pm$ 0.421 \\
8996.64154 & 0.89 & 25.589 $\pm$ 0.472 \\
9052.50644 & 0.39 & -84.507 $\pm$ 0.377 \\
9052.52226 & 0.41 & -67.123 $\pm$ 0.357 \\
9052.53807 & 0.43 & -60.867 $\pm$ 0.511 \\
9053.50377 & 0.54$^{*}$ & 3.202 $\pm$ 0.513 \\
9053.51966 & 0.56$^{*}$ & 13.055 $\pm$ 0.314 \\
9053.53540 & 0.58 & 18.949 $\pm$ 0.331 \\
9053.55098 & 0.60 & 27.319 $\pm$ 0.327 \\
9054.48047 & 0.67 & 59.870 $\pm$ 1.516 \\
9054.50819 & 0.71 & 65.697 $\pm$ 1.313 \\
9054.52371 & 0.72 & 66.848 $\pm$ 1.321\\
9054.55120 & 0.76 & 62.872 $\pm$ 1.490 \\
9054.56680 & 0.77 & 61.067 $\pm$ 1.358 \\
9055.53483 & 0.89 & 30.682 $\pm$ 0.500\\
9055.55030 & 0.91 & 18.770 $\pm$ 0.571 \\
9055.56834 & 0.93 & 13.521 $\pm$ 0.536 \\
9055.58374 & 0.95 & 8.488 $\pm$ 0.456\\
9055.59933 & 0.96 & 0.392 $\pm$ 0.492\\
9056.44893 & 0.94 & 11.107 $\pm$ 0.552\\
9058.47553 & 0.28 & -111.006 $\pm$ 1.289\\
9059.46862 & 0.43 & -56.888 $\pm$ 0.607\\
9060.45183 & 0.57 & 13.703 $\pm$ 0.637\\
9090.47989 & 0.24 & -111.172 $\pm$ 0.503\\
9093.38956 & 0.60 & 33.861 $\pm$ 1.110\\
9093.40534 & 0.62 & 40.637 $\pm$ 1.308\\
9093.42083 & 0.63 & 46.740 $\pm$ 0.868\\
9093.43644 & 0.66 & 54.564 $\pm$ 0.921\\
9094.40102 & 0.77 & 67.568 $\pm$ 1.313\\
9094.41671 & 0.78 & 62.270 $\pm$ 1.219\\
9094.43230 & 0.80 & 60.672 $\pm$ 1.153\\
9096.36986 & 0.04 &-49.920 $\pm$ 0.682\\
9096.38560 & 0.06 & -55.208 $\pm$ 0.387\\
9096.40135 & 0.08 &-61.628 $\pm$ 0.546\\
  \noalign{\smallskip}\hline
\end{tabular}
\end{center}  
\end{table}

\begin{table}
\begin{center}
\caption[]{The minima times derived from the TESS data and our photometric observation in V and R filter. I and II repsresent the primary and secondary minimums. For the V,R photomteric observation a median minima time is given.}\label{minima_table}
 \begin{tabular}{lcc}
  \hline\noalign{\smallskip}
  Time of minima & Filter  & Minima type \\
  HJD (2450000+) &   \\
  \hline\noalign{\smallskip}
8983.74370 $\pm$ 0.00003 & TESS & I \\
8985.47575 $\pm$ 0.00002 & TESS & I \\
9008.85843 $\pm$ 0.00003 & TESS & I \\
8984.17794 $\pm$ 0.00011 & TESS & II \\
8985.04381 $\pm$ 0.00020 & TESS & II \\
9003.23004 $\pm$ 0.00008 & TESS & II \\
9010.59182 $\pm$ 0.00003 & TESS & I \\
9025.31420 $\pm$ 0.00004 & TESS & I \\
9032.24248 $\pm$ 0.00003 & TESS & I \\
9011.89158 $\pm$ 0.00010 & TESS & II \\
9025.74788 $\pm$ 0.00009 & TESS & II \\
9032.67621 $\pm$ 0.00009 & TESS & II \\
9405.49803 $\pm$ 0.00029 & VR   &  I\\
  \noalign{\smallskip}\hline
\end{tabular}
\end{center}  
\end{table}

\end{document}